\documentclass[a4paper]{jpconf}
\usepackage{graphicx}

\begin{document}

\title{The contribution of pseudoscalar and axial-vector mesons to hyperfine
structure of muonic hydrogen}

\author{A.~E.~Dorokhov}

\address{Joint Institute of Nuclear Research, BLTP,
141980, Moscow region, Dubna, Russia}

\ead{dorokhov@theor.jinr.ru}

\author{N.~I.~Kochelev}

\address{Institute of Modern Physics of Chinese Academy of Sciences, 730000, Lanzhou, China\\
Joint Institute of Nuclear Research, BLTP,
141980, Moscow region, Dubna, Russia}

\ead{nikkochelev@mail.ru}

\author{A.~P.~Martynenko}

\address{Samara University, 443086, Samara, Russia}

\ead{a.p.martynenko@samsu.ru}

\author{F.A.~Martynenko}

\address{Samara University, 443086, Samara, Russia}

\ead{f.a.martynenko@gmail.com}

\author{A.~E.~ Radzhabov}

\address{Institute of Modern Physics of Chinese Academy of Sciences, 730000, Lanzhou, China\\
Matrosov Institute for System Dynamics and Control Theory SB RAS, 664033, Irkutsk, Russia}

\ead{aradzh@icc.ru}

\author{R.N. Faustov}

\address{Institute of Informatics in Education, FRC CSC RAS, 119333, Moscow,
Russia}

\begin{abstract}
In the framework of the quasipotential method in quantum electrodynamics we
calculate the contribution of light pseudoscalar (PS) and axial-vector (AV)
mesons to the interaction operator of a muon and a proton in muonic hydrogen
atom. The coupling of mesons with the muon is via two-photon intermediate
state. The parametrization of the transition form factor of two photons into
PS and AV mesons, based on the experimental data on the transition form
factors and QCD asymptotics, is used. Numerical estimates of the
contributions to the hyperfine structure of the spectrum of the S and P
levels are presented. It is shown that such contribution to the hyperfine
splitting in muonic hydrogen is rather important for a comparison
with precise experimental data.
\end{abstract}

\section{Proton radius puzzle}

Precise investigation of the Lamb shift (LS) and hyperfine structure (HFS)
of light muonic atoms is a fundamental problem for testing the Standard
model and establishing the exact values of its parameters, as well as
searching for effects of new physics. Recently, the CREMA (Charge Radius
Experiments with Muonic Atoms) Collaboration from PSI by using the laser
spectroscopy method measured with unprecedented accuracy the transition
frequencies between the 2P and 2S states in muonic hydrogen ($\mu p$) \cite%
{crema1,crema2}%
\begin{eqnarray}
\nu _{t}\left( 2\mathrm{P}_{3/2}^{F=2}-2\mathrm{S}_{1/2}^{F=1}\right)
&=&49881.35(65)~\mathrm{GHz,\qquad }h\nu _{t}=\Delta E\left( 2\mathrm{P}%
_{3/2}^{F=2}-2\mathrm{S}_{1/2}^{F=1}\right)   \label{nut} \\
\nu _{s}\left( 2\mathrm{P}_{3/2}^{F=1}-2\mathrm{S}_{1/2}^{F=0}\right)
&=&54611.16(1.05)~\mathrm{GHz,\qquad }h\nu _{s}=\Delta E\left( 2\mathrm{P}%
_{3/2}^{F=1}-2\mathrm{S}_{1/2}^{F=0}\right) .  \label{nus}
\end{eqnarray}%
From these two transition measurements, both the Lamb shift $\Delta E_{LS}$
and the 2S-HFS $\Delta E_{HFS}$ was determined independently with the result
\cite{crema2}%
\begin{eqnarray}
&&\Delta E_{LS}^{\exp }\left( 2\mathrm{P}_{1/2}-2\mathrm{S}_{1/2}\right)
=202.3706(23)~\mathrm{meV,}  \label{DELexp} \\
&&\Delta E_{HFS}^{\exp }\left( 2\mathrm{S}_{1/2}^{F=1}-2\mathrm{S}%
_{1/2}^{F=0}\right) =22.8089(51)~\mathrm{meV.}  \label{DEHFSexp}
\end{eqnarray}%
From theory side these quantities calculated in the frame work of
bound-state QED are expressed in terms of the charge $\mathrm{r}_{E}^{2}$
and Zemach $\mathrm{r}_{Z}$ radii of the proton as \cite{crema3}%
\begin{eqnarray}
&&\Delta E_{\mathrm{LS}}^{\mathrm{th}}=206.0668(25)-5.2275(10)~\mathrm{r}%
_{E}^{2}\,~\mathrm{meV,}  \label{DELth} \\
&&\Delta E_{\mathrm{HFS}}^{\mathrm{th}}=22.9843(30)-0.1621(10)~\mathrm{r}%
_{Z}~\mathrm{meV}.  \label{DEHFSth}
\end{eqnarray}%
Remind that the charge RMS radius $\mathrm{r}_{E}$ is defined via the
normalized proton charge distribution $\rho _{\mathrm{E}},$ while the Zemach
radius is correlated with the proton magnetic moment distribution $\rho _{%
\mathrm{M}}$ as%
\begin{equation}
\mathrm{r}_{\mathrm{E}}^{2}=\int d^{3}rr^{2}\rho _{\mathrm{E}}\left(
r\right) ,\qquad \mathrm{r}_{Z}=\int d^{3}r\int d^{3}r^{\prime }r^{\prime
}\rho _{\mathrm{E}}\left( r\right) \rho _{\mathrm{M}}\left( r-r^{\prime
}\right) .  \label{RhoEM}
\end{equation}%
The first term on the right side of (\ref{DELth}) accounts for radiative,
relativistic, and recoil effects, while the first term on the right side of (%
\ref{DEHFSth}) is the Fermi energy arising from the interaction between the
muon and the proton magnetic moment and different corrections to it. It is
also important, that like for the problem of the anomalous magnetic moments
of leptons \cite{dorokhov3,Dorokhov:2014iva}, the coefficients in front of
the radius terms in (\ref{DELth}), (\ref{DEHFSth}) are much stronger
enhanced for the muonic system relative to the electronic system, and thus
much more sensitive in extraction of the radii parameters from the
experimental data.

The comparison of (\ref{DELth}),(\ref{DEHFSth}) with (\ref{DELexp}),(\ref%
{DEHFSexp}) provides \cite{crema2,crema3}%
\begin{equation}
\mathrm{r}_{\mathrm{E}}^{\mathrm{CREMA}}=0.84087(39)~~\mathrm{fm,\qquad r}_{%
\mathrm{Z}}^{\mathrm{CREMA}}=1.082(37)~~\mathrm{fm.}  \label{rEZcrema}
\end{equation}%
This should be compared with the values recommended by CODATA \cite%
{Mohr:2012tt} and based on the electron-proton scattering and electronic
hydrogen spectroscopy%
\begin{equation}
\mathrm{r}_{\mathrm{E}}^{\mathrm{CODATA}}=0.8751(61)~~\mathrm{fm.\qquad }
\label{rEZcodata}
\end{equation}

Thus we see two basic results of the laser spectroscopy experiment for $\mu
p$. First, from the $\mu p$ spectroscopy the value of $\mathrm{r}_{\mathrm{E}%
}$ is determined with precision 10 times higher than from electronic data.
Second, and more striking, that there is the large discrepancy between $%
\mathrm{r}_{\mathrm{E}}^{\mathrm{CREMA}}$ and $\mathrm{r}_{\mathrm{E}}^{%
\mathrm{CODATA}}$ at the level of 5.6 $\sigma $ (or the muonic hydrogen
value is 4\% smaller than the CODATA value$)$. This is so-called the \textbf{%
proton size puzzle}, not explained up to now! Later on, the similar problem
was detected for the deutron radius \cite{crema4}.

At present, several experimental groups plan to measure HFS of various
muonic atoms with more high precision \cite{ma-2017,adamczak-2017,pohl-2017}%
. This will make it possible to better understand the existing "puzzle" of
the proton charge radius, to check the Standard Model with greater accuracy
and, possibly, to reveal the source of previously unaccounted interactions
between the particles forming the bound state in QED. One of the ways of
overcoming the crisis situation arises from a deeper theoretical analysis of
the fine and hyperfine structure of muonic atom spectrum, with the
verification of previously calculated contributions and the more accurate
construction of the particle interaction operator in quantum field theory,
the calculation of new corrections whose value for muonic atoms can increase
substantially in comparison with electronic atoms. The expected results will
allow to get also a new very important information about the forces which
are responsible for the structure of atoms.

\section{Light meson exchange contributions to HFS of $\protect\mu p$}

\begin{figure}[th]
\centerline{
\includegraphics[scale=1.0]{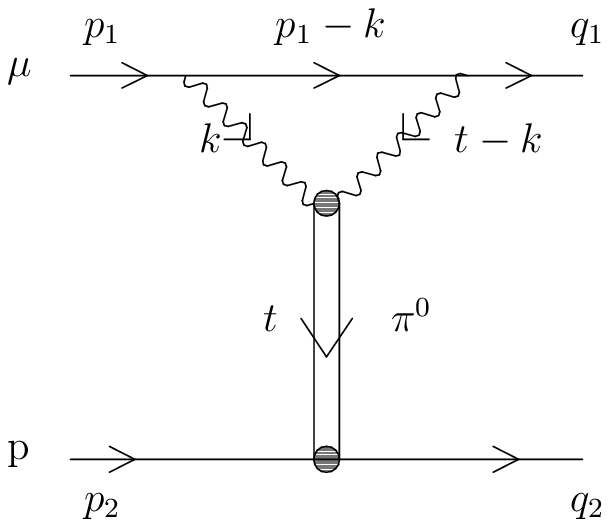}
\includegraphics[scale=1.0]{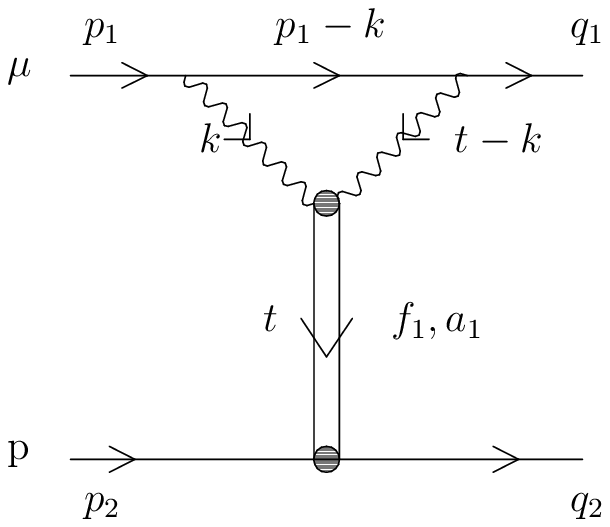}
}
\caption{Muon-proton interaction induced by mesonic exchange.}
\label{fig1}
\end{figure}

From the theory side it is urgently needed to study the possible effects of
exchanges between muon and proton which can contribute to HFS of $\mu p$.
Some of such effects was considered in recent papers \cite%
{apmlet,Dorokhov:2017nzk,pang,pascalutsa,kou}. Below, we discuss the effects
of exchanges between muon and proton which can contribute to HFS of $\mu p$
coming from the light pseudoscalar (PS) and axial-vector (AV) meson
exchanges between muon and proton induced by meson coupling to muon through
two photons (see Fig.~\ref{fig1}). \

The leading contribution to HFS of $\mu p$ is coming from one-photon
exchange and has the following form \cite{apm2004,apm1999,pra2016}:
\begin{eqnarray}
\Delta V_{B}^{hfs} &=&\frac{8\pi \alpha \mu _{p}}{3m_{\mu }m_{p}}(\mathbf{S}%
_{p}\mathbf{S}_{\mu })\delta (\mathbf{r})-\frac{\alpha \mu _{p}(1+a_{\mu })}{%
m_{\mu }m_{p}r^{3}}\left[ (\mathbf{S}_{p}\mathbf{S}_{\mu })-3(\mathbf{S}_{p}%
\mathbf{n})(\mathbf{S}_{p}\mathbf{n})\right]   \label{eq:3} \\
&&+\frac{\alpha \mu _{p}}{m_{\mu }m_{p}r^{3}}\left[ 1+\frac{m_{\mu }}{m_{p}}-%
\frac{m_{\mu }}{2m_{p}\mu _{p}}\right] (\mathbf{L}\mathbf{S}_{p})  \nonumber
\end{eqnarray}%
where $m_{\mu }$, $\mathbf{S}_{\mu }$ and $m_{p}$, $\mathbf{S}_{p}$ are
masses and spins of muon and proton, correspondingly, $\mu _{p}$ is the
proton magnetic moment. The potential (\ref{eq:3}) gives the main
contribution of order $\alpha ^{4}$ to the HFS of muonic atom. Precision
calculation of the HFS of the spectrum, which is necessary for comparison
with experimental data, requires the consideration of various corrections
accounting for the vacuum polarization, nuclear structure and recoil, and
relativistic effects \cite{apm2004,kp1,egs,borie,apm2002,apm2017}.

We calculate further the contribution to HFS coming from the pion \footnote{The contribution of the $\eta$ and $\eta'$ mesons is negligible \cite{apmlet}} and
axial-vector $f_{1}(1285)$, $a_{1}(1260)$ and $f_{1}(1420)$ meson exchanges
shown in Fig.~\ref{fig1}. The effective vertices of the interaction of the
PS and AV mesons and virtual photons can be expressed in terms of the
transition form factors as follows:
\begin{eqnarray}
&&V^{\mu \nu }(k_{1},k_{2})=i\varepsilon ^{\mu \nu \alpha \beta }k_{1\alpha
}k_{2\beta }\frac{\alpha }{\pi F_{\pi }}F_{\pi ^{0}\gamma ^{\ast }\gamma
^{\ast }}(k_{1}^{2},k_{2}^{2}),  \label{eq:4} \\
&&T^{\mu \nu \alpha }=8\pi i\alpha \varepsilon _{\mu \nu \alpha \tau
}k^{\tau }k^{2}F_{AV\gamma ^{\ast }\gamma ^{\ast
}}^{(0)}(k_{1}^{2},k_{2}^{2}),  \label{Tmunu}
\end{eqnarray}%
where $k_{1}$, $k_{2}$ are four-momenta of virtual photons. For small values
of the relative momenta of particles in the initial and final states and
small value of transfer momentum $t$ between muon and proton, the transition
amplitude takes a simple form
\begin{equation}
T^{\mu \nu \alpha }=8\pi i\alpha \varepsilon _{\mu \nu \alpha \tau }k^{\tau
}k^{2}F_{AV\gamma ^{\ast }\gamma ^{\ast }}^{(0)}(t^{2},k^{2},k^{2}),
\label{eq:a1a}
\end{equation}%
where $k=k_{1}=-k_{2}$.

The final result for the HFS potential is equal to
\begin{eqnarray}
&&\Delta V_{PS}^{hfs}(\mathbf{t})=\frac{\alpha ^{2}}{6\pi ^{2}}%
\frac{g_{p}}{m_{p}F_{\pi }}\frac{\mathbf{t}^{2}}{\mathbf{t}^{2}+m_{\pi }^{2}}%
\mathcal{A}(\mathbf{t}^{2})
,  \label{DVPS} \\
&&\Delta V_{AV}^{hfs}(\mathbf{t})=-\frac{32\alpha ^{2}g_{AVPP}F_{AV\gamma\gamma}(0,0)}{3\pi^2(\mathbf{t}^{2}+M_{A}^{2})}%
I(\frac{m_\mu}{\Lambda_A}).  \label{eq:a4}
\end{eqnarray}%
where
\begin{eqnarray}
&&\mathcal{A}(\mathbf{t}^{2})=\frac{2}{\pi^2\mathbf{t}^{2}}\int \frac{id^{4}k[t^{2}k^{2}-(tk)^{2}]}{%
k^{2}(k-t)^{2}(k^{2}-2kp_{1})}F_{PS\gamma ^{\ast }\gamma ^{\ast
}}(k^{2},(k-t)^{2}),\label{A}\\
&&I(\frac{m_\mu}{\Lambda_A})= \int \frac{%
id^{4}k(2k^{2}+k_{0}^{2})}{k^{2}(k^{2}-2m_{\mu }k_{0})}F_{AV\gamma ^{\ast
}\gamma ^{\ast }}(k^{2},k^{2}).\label{I}
\end{eqnarray}
The integral $\mathcal{A}$ in (\ref{A}) is well studied in
connection to the problem of interpretation \cite{dorokhov2,Dorokhov:2009xs}
of the KTeV (FermiLab) data on the pion decay into $e^{+}e^{-}$ pair. In
order to fix the transition form factors in the most model independent way
we used corresponding data from CLEO \cite{Gronberg:1997fj} and L3 \cite%
{L3C,L3Ca,aihara} collaborations for the PS and AV mesons, respectively (see
for details \cite{apmlet,Dorokhov:2017nzk})\footnote{In recent paper \cite{Osipov:2017ray}within NJL model the  estimation for form factor $F^{0(NJL)}_{f_1(1285),\gamma^*\gamma^*}\approx 0.276 GeV^{-2} $
was obtained. This value is in the agreement with L3 data which   was used in our paper \cite{Dorokhov:2017nzk} (see Table 1.)}. We would like to point out,
that one can expect the important contribution of the AV exchange to spin
dependent part of muon-proton interaction because the exchange particle has
the spin one. Furthermore, it is also well known that in the channel with
quantum number $1^{++}$ axial anomaly effects can play an important role
and, in particularly, these effects might be considered as a cornerstone to
solve so-called "proton spin crisis" \cite{Dorokhov:1993ym,Anselmino:1994gn}%
. The other phenomenological input, the meson-nucleon couplings, were
determined by using the Regge approach analysis of the deep-inelastic
scattering, $f_{1}$ and $a_{1}$ trajectories contributing to the
polarization of quarks in the nucleon \cite{kochelev,kochelev1}.

Calculating the matrix elements with wave functions of 1S , 2S and 2P 1/2
states, we obtain the corresponding contributions to the HFS spectrum%
\begin{equation}
\Delta E^{hfs}(1S)=\frac{\mu ^{3}\alpha ^{5}g_{A}}{6F_{\pi }^{2}\pi ^{3}}%
\Biggl\{\mathcal{A}(0)\frac{4W(1+\frac{W}{m_{\pi }})}{m_{\pi }(1+\frac{2W}{%
m_{\pi }})^{2}}-\frac{1}{\pi }\int_{0}^{\infty }\frac{ds}{s}Im\mathcal{A}%
(s)\times  \label{eq:11}
\end{equation}%
\[
\left[ 1+\frac{1}{4W^{2}(s-m_{\pi }^{2})}\left( \frac{m_{\pi }^{4}}{(1+\frac{%
m_{\pi }}{2W})^{2}}-\frac{s^{2}}{(1+\frac{\sqrt{s}}{2W})^{2}}\right) \right] %
\Biggr\},
\]%
\begin{equation}
\Delta E^{hfs}(2S)=\frac{\mu ^{3}\alpha ^{5}g_{A}}{48F_{\pi }^{2}\pi ^{3}}%
\Biggl\{\mathcal{A}(0)\frac{W(8+11\frac{W}{m_{\pi }}+8\frac{W^{2}}{m_{\pi
}^{2}}+2\frac{W^{3}}{m_{\pi }^{3}})}{2m_{\pi }(1+\frac{W}{m_{\pi }})^{4}}-%
\frac{1}{\pi }\int_{0}^{\infty }\frac{ds}{s}Im\mathcal{A}(s)\times
\label{eq:12}
\end{equation}%
\[
\left[ 1+\frac{1}{(s-m_{\pi }^{2})}\left( \frac{m_{\pi }^{2}(2+\frac{W^{2}}{%
m_{\pi }^{2}})}{2(1+\frac{W}{m_{\pi }})^{4}}-\frac{s(2+\frac{W^{2}}{s})}{2(1+%
\frac{W}{\sqrt{s}})^{4}}\right) \right] \Biggr\},
\]
\begin{equation}
\Delta E_{AV}^{hfs}(1S)=\frac{32\alpha ^{5}\mu ^{3}g_{AVPP}F_{AV\gamma
^{\ast }\gamma ^{\ast }}^{(0)}(0,0,0)}{3M_{A}^{2}\pi ^{3}\Bigl(1+\frac{2W}{%
M_{A}}\Bigr)^{2}}I\left( \frac{m_{\mu }}{\Lambda _{A}}\right) ,
\label{eq:a6}
\end{equation}%
\begin{equation}
\Delta E_{AV}^{hfs}(2S)=\frac{2\alpha ^{5}\mu ^{3}g_{AVPP}F_{AV\gamma ^{\ast
}\gamma ^{\ast }}^{(0)}(0,0,0)\left( 2+\frac{W^{2}}{M_{A}^{2}}\right) }{%
3M_{A}^{2}\pi ^{3}\Bigl(1+\frac{W}{M_{A}}\Bigr)^{4}}I\left( \frac{m_{\mu }}{%
\Lambda _{A}}\right) ,  \label{eq:a6a}
\end{equation}%
where $W=\mu \alpha $ and $\mu $ is reduced mass.

\begin{table}[h]
\caption{Pion and axial-vector mesons exchanges contribution to HFS of muonic hydrogen.}
\label{tb1}
\bigskip
\begin{tabular}{|c|c|c|c|c|c|}
\hline
mesons & $I^G(J^{PC})$ & $\Lambda_A$ & $F^{(0)}_{AV\gamma^\ast\gamma^%
\ast}(0,0)$ & $\Delta E^{hfs}(1S)$ & $\Delta E^{hfs}(2S)$ \\
&  & in MeV & in $GeV^{-2}$ & in meV & in meV \\ \hline
$f_1(1285)$ & $0^+(1^{++})$ & 1040 & 0.266 & $-0.0093\pm 0.0033$ & $%
-0.0012\pm 0.0004$ \\ \hline
$a_1(1260)$ & $1^-(1^{++})$ & 1040 & 0.591 & $-0.0437\pm 0.0175$ & $%
-0.0055\pm 0.0022$ \\ \hline
$f_1(1420)$ & $0^+(1^{++})$ & 926 & 0.193 & $-0.0013\pm 0.0008$ & $%
-0.0002\pm 0.0001$ \\ \hline
$\pi^0$ & $1^-(0^{-+})$ & 776 & \phantom{.} & $-0.0017\pm 0.0001$ & $-0.0002\pm 0.00002$
\\\hline\hline
Sum & \phantom{.} & \phantom{.} & \phantom{.} & $-0.0560\pm 0.0178$ & $-0.0071\pm 0.0024$
\\ \hline\hline
\end{tabular}%
\end{table}

In Table ~\ref{tb1} our results for contribution of the pion and AV mesons exchanges
to HFS are presented. For the case of $2S$ state the summary contribution
from pion and AV meson exchanges is equal to (-0.0071) meV, which is quite important to obtain the total value of HFS with high precision.

\section{Conclusion}

A new important contribution to the muon-nucleon interaction is discovered.
It is coming from effective pion and AV mesons exchanges induced by anomalous
meson vertices with two photon state. The contribution of this exchange to
the HFS of $\mu p$ is calculated in framework of the quasipotential method
in QED and the use of the technique of projection operators on the states of
two particles with a definite spin. It is shown that this contribution is
rather large and should be taking into account for the interpretation of the
new data on HFS in this atom.

As has been mentioned the CREMA Collaboration measured two transition
frequencies in muonic hydrogen for the 2S triplet state $%
(2P_{3/2}^{F=2}-2S_{1/2}^{F=1})$ and for 2S singlet state $%
(2P_{3/2}^{F=1}-2S_{1/2}^{F=0})$ \cite{crema2}. From these measurements it
is possible to extract the value of HFS for $2S$ level. Obtained value $(\ref%
{DEHFSexp})$ allows to get the value of the Zemach radius (\ref{rEZcrema})
with accuracy $3.4~\%$ with help of relation (\ref{DEHFSth}). This is in the
agreement with another numerical values $r_{Z}=1.086(12)$ fm \cite{p1}, $%
r_{Z}=1.045(4)$ fm \cite{p2}, $r_{Z}=1.047(16)$ fm \cite{p3}, $%
r_{Z}=1.037(16)$ fm \cite{p4} obtained from electron-proton scattering and
from H and muonium spectroscopy. We should emphasize that the changing of
the theoretical value of HFS on 0.001 meV leads to the changing of the
Zemach radius on 0.006 fm. Therefore, our contribution coming from pion and AV meson
exchange leads to the new value of this radius
\begin{equation}
r_{Z}=1.040(37)~\mathrm{fm,}  \label{rzour}
\end{equation}%
which is smaller in the comparison with most listed results but still agree
with them within errorbars.

The work is supported by Russian Science Foundation (grant No. RSF 15-12-10009) (A.E.D.),
the Chinese Academy of Sciences visiting professorship for senior international scientists
(grants No. 2013T2J0011) (N.I.K.) and President's international fellowship initiative
(Grant No. 2017VMA0045) (A.E.R.), Russian Foundation for Basic Research (grant No. 16-02-00554) (A.P.M., F.A.M.).

\end{document}